\documentclass[twocolumn,prl,aps,showpacs,floatfix,superscriptaddress]{revtex4}

\usepackage{graphicx}

\begin{document}

\newcommand{\SrTwoLayer}{Sr$_3$Ru$_2$O$_7$} 
\newcommand{\SrOneLayer}{Sr$_2$RuO$_4$} 
\newcommand{\SrInftyLayer}{SrRuO$_3$} 
\renewcommand{\floatpagefraction}{0.5}

\title{Critical end-point and metamagnetic 
quantum criticality in Sr$_3$Ru$_2$O$_7$}
  
\author{May Chiao}

\thanks{Present address: Laboratorium f\"ur Festk\"orperphysik, 
Eidgen\"ossische Technische Hochschule,
ETH-Z\"urich, CH-8093 Z\"urich, Switzerland.}

\affiliation{Cavendish Laboratory, University of Cambridge, Madingley
  Road, Cambridge CB3 0HE, United Kingdom}

\author{C. Pfleiderer}

\affiliation{Physikalisches Institut, Universit\"at Karlsruhe,
  Wolfgang-Gaede-Strasse 1, D-76128 Karlsruhe, Germany}
  
\author{R. Daou}

\affiliation{Cavendish Laboratory, University of Cambridge, Madingley
  Road, Cambridge CB3 0HE, United Kingdom}
  
\author{A. McCollam}

\affiliation{Cavendish Laboratory, University of Cambridge, Madingley
  Road, Cambridge CB3 0HE, United Kingdom}

\author{S. R. Julian}

\affiliation{Cavendish Laboratory, University of Cambridge, Madingley
  Road, Cambridge CB3 0HE, United Kingdom}

\author{G. G. Lonzarich}

\affiliation{Cavendish Laboratory, University of Cambridge, Madingley
  Road, Cambridge CB3 0HE, United Kingdom}

\author{R. S. Perry}

\affiliation{School of Physics and Astronomy, University of Birmingham,
Edgbaston, Birmingham BI5 2TT, United Kingdom}

\author{A. P. Mackenzie}

\affiliation{School of Physics and Astronomy, University of 
St. Andrews, North Haugh, St. Andrews, Fife KY16 9SS, United Kingdom}

\author{Y. Maeno}

\affiliation{Department of Physics, Kyoto University, Kyoto 606-8502, Japan}

\affiliation{CREST, Japan Science and Technology Corporation,
  Kawaguchi, Saitama 332-0012, Japan}

\date{\today}

\begin{abstract}
We report a metamagnetic critical point at ($B^*$,$T^*$) $\simeq$
(5.1\,T, 1.1\,K) for magnetic fields applied perpendicular to the
(tetragonal) $c$-axis of \SrTwoLayer.
First-order behaviour well below $T^*$ indicates that $(B^*,T^*)$
marks a critical end-point that terminates a line of first-order
metamagnetic transitions.
The absence of first-order behaviour in the metamagnetic transition
with $B \parallel c$ confirms that the non-Fermi liquid behaviour for
$B \parallel c$ is underpinned by a metamagnetic quantum critical
end-point for which $T^* \to 0$.
Scaling behaviour of the resistivity under hydrostatic pressure yields
the surprising result that although $B^*$ increases rapidly with
pressure, $T^*$ has only a weak pressure dependence.  
\end{abstract}



\maketitle

The intense exploration of the perovskite ruthenates over the past few
years has revealed a wide range of challenges to our understanding of
the metallic state.
This family includes the unconventional superconductor \SrOneLayer\
\cite{Maeno} and the itinerant ferromagnet \SrInftyLayer\
\cite{Allen,Klein}; Ca-doping \SrOneLayer\ at the Sr site even
provides access to a metal-insulator transition \cite{Braden}.
We recently reported evidence of non-Fermi liquid behaviour in
\SrTwoLayer\ in the vicinity of a metamagnetic transition which occurs
at fields $B_M(T)$ which depend on field orientation, with
$B_{M\perp}(0)$ = 5.1\,T for $B\perp c$, and $B_{M\parallel}(0)$ =
7.8\,T for $B||c$ \cite{Perry,Grigera}.
There are two regimes of non-Fermi liquid behaviour.
Above 1\,K the metamagnetic transition broadens rapidly with
temperature indicating a fall in the susceptibility for $B \sim B_M$
and there is a quasi-linear temperature dependence of the resistivity
over a large region of the $(B,T)$ plane centered on $B_M$; also, in
measurements with $B\parallel c$, a logarithmic divergence has been
seen in the electronic specific heat, $C(T)/T \sim -\ln T$
\cite{Perry}.
Below 1\,K in contrast, non-Fermi liquid behaviour is confined to $B
\sim B_{M\parallel}$; for other values and orientations of $B$ the
susceptibility and resistivity cross-over to conventional Fermi-liquid
behaviour.
The resistivity in particular shows $T^2$ behaviour, but the range of
temperature in which this is observed collapses as $B\rightarrow
B_{M\parallel}$ while the coefficient $A$ in $\rho(T) = \rho_0 + A
T^2$ becomes divergent.
Concurrently, for $T \rightarrow 0$ (and $B \parallel c$) a novel
manifestation of non-Fermi liquid behaviour appears as a power law in
the resistivity that is {\em higher} than $T^2$, being closer to $T^3$
\cite{Grigera}.
It has been speculated that the $T^{3}$ behaviour signals the onset of a
qualitatively new state, driven by divergent fluctuations and
susceptibility associated with a so-called metamagnetic quantum
critical end-point \cite{Grigera}.

Historically, metamagnetism refers to magnetic field induced phase
transitions of local moment antiferromagnetic insulators.
Instead, the itinerant metamagnetism addressed in this Letter is
thought to arise from a magnetic field induced spin splitting of the
Fermi surface.
Metamagnetism of this kind has been detected in numerous metallic
magnets; in some systems the rapid rise in the magnetisation $M$
is continuous (\lq metamagnetic-like' behaviour \cite{Flouquet95,Puech88})
but in others the metamagnetic transition can be a true phase transition with
a discontinuous jump in $M$ as a function of $B$ \cite{Goto_review}.
In a true metamagnetic transition the amplitude of the metamagnetic jump
$\Delta M$ decreases as $T$ increases,
vanishing at a critical point $(B^*,T^*)$ in the $(B,T)$ plane.
Above $T^{*}$ one sees crossover behaviour.
The termination of the line of first-order transitions at an isolated 
critical end-point is analogous to the liquid--vapour co-existence curve.
If the critical end-point were to occur at $T^* = 0$ one would have
a metamagnetic quantum critical end-point.

Non-Fermi liquid behaviour is often seen in systems with metamagnetic or
metamagnetic-like transitions.
For instance, UCoAl \cite{Andreev85} has a first order metamagnetic
transition at $B\sim$0.65\,T, and a critical point at $(B^*,T^*)
\sim$(0.8\,T,13\,K) \cite{Mushnikov99}.
However, the reported non-Fermi liquid behaviour in the resistivity as
$T\to 0$ is assumed to arise from the proximity to a zero field
ferromagnetic quantum critical point, rather than a metamagnetic one
\cite{Kolomiets99}.
In MnSi, though, the vicinity to itinerant metamagnetism at high
pressures may be responsible for the observed non-Fermi liquid
behaviour over a very large region of $T$ and $p$, but the quantum
critical end-point does not appear to play a special role \cite{MnSi}.
Regarding the heavy fermion metamagnetic systems, such as
CeRu$_2$Si$_2$ \cite{Flouquet95,Puech88} and UPt$_3$
\cite{Kim00,Heuser00,Aoki98}, non-Fermi liquid properties have also
been observed, but none of the heavy fermion systems studied to date
have shown a first-order jump in $M$ for any combination of
parameters.
Thus the relevance of fluctuations associated with a quantum critical
end-point cannot be unambiguously demonstrated.

In this Letter we report the presence of a line of first order metamagnetic
phase transitions in \SrTwoLayer\ terminating in a critical end-point
for $B \perp c$. 
The absence of such first-order behaviour in the metamagnetic
transition with $B\parallel c$ implies that $T^*$ vanishes at some
intermediate angle, thus supporting the suggestion that a quantum
critical end-point underlies the non-Fermi liquid behavior for $B
\parallel c$ \cite{Grigera}.  
Moreover we investigate how the critical point evolves under pressure, 
showing that although $B^*$ rises rapidly with pressure, $T^*$ is 
comparatively unaffected. 
This unexpected pressure dependence of the critical end-point,
inferred from the resistivity, shows that quantum criticality in
\SrTwoLayer\ is not consistent with the standard model of itinerant 
metamagnetism \cite{Yamada93}.

The single crystals of Sr$_3$Ru$_2$O$_7$ examined here were grown from
high purity starting materials in an infrared image furnace in Kyoto.
The high sample quality was confirmed by measurements of the d.c.
magnetisation and resistivity ratio before and after pressurisation in
a miniature Cu:Be clamp cell.
The ambient pressure properties of the samples studied here correspond
to those observed and reported for around 50 other crystals from the
same batch \cite{Perry}.

A.c.\ susceptibility measurements were carried out on a 1\,mm$^{3}$
sample in an 18\,T cryomagnetic facility with a base temperature of
6\,mK.
A small modulation field of less than 0.01\,T was applied at a frequency 
of 77\,Hz.
Au wires were Ag-epoxied onto the samples and then soldered to 
Cu-clad NbTi leads inside the pressure cell.
A 1:4 methanol-ethanol mixture served as pressure transmitting medium
which ensured that the pressure was highly isotropic at low $T$.  
The low temperature resistivity was measured in a commercial 12\,T
variable temperature $^4$He cryostat in the range 2\,K to 300\,K using a
low frequency four terminal a.c. technique.

In Fig.~1 we show susceptibility curves $\chi \equiv (\partial
M/\partial H)_{T,p}$ vs $B$, applied perpendicular to $c$, at
temperatures below 2\,K.
The peak in the susceptibility at $B_{M\perp}$ grows with decreasing
temperature down to 1.1\,K, but thereafter the peak falls sharply in
size and becomes cusp-like.
A cusp in $\chi (B)$ indicates a discontinuity in $M(B)$, since the
continuous change in magnetisation across the transition is given by
the integral of $\chi$ with respect to $B$.
Torque magnetometry also shows a clear jump in the torque at the
transition \cite{Alix}.
These results unambiguously show that a first order jump develops in
$M$ below about 1\,K in this sample.  
Thus we interpret the peak in $\chi$ at (5.1\,T, 1.1\,K) to be 
some kind of critical point.  
The inset of Fig.~1 shows the position of the peak in the $(B,T)$
plane, with a solid line representing the first-order regime.
Nevertheless, the magnitude of the peak, given in absolute units in Fig.~1,
does not diverge as expected from a naive Landau-Ginzburg free
energy for a metamagnetic transition \cite{Yamada93}, nor from a more
sophisticated treatment more appropriate to a quantum critical
metamagnetic transition \cite{Millis02,GGL}.
The simplest explanation we can suggest is that slight inhomogeneities
in the sample smear out the transition, and indeed $T^{*}$ has shown
some variation between samples \cite{Grigera-next} though we believe that
this may reflect in-plane anisotropy of the metamagnetic transition.

We turn next to the basal plane resistivity, used as a probe of the
electronic behaviour.  Shown in Fig.~2 is the normalised, isothermal
magnetoresistance $\rho/\rho (B=0)$ measured at 1.5\,K at several
pressures up to 10\,kbar, for both $B~{||}~c$ and $B{\perp}c$.
At the lowest $T$ a double peak structure may be resolved at the fields 
$B_{M\perp 1}$ and $B_{M\perp 2}$ for $B \perp c$.
Together with $B_{M\parallel}$ (for $B~||~c$) there are three
identifiable anomalies.
An extrapolation of these three transition fields to negative
pressures, shown in Fig.~3, reveals that they all extrapolate to
$B_{M}$=0 at the {\em same} negative pressure $p_{c\perp} = p_{c ||} =
p_c \approx -14$ kbar.
This suggests that at a hypothetical negative pressure $p_{c}$ there
is a zero-field ferromagnetic transition.
The existence of a zero-field quantum critical point at negative 
pressure is reinforced by the fall in $A$, the $T^2$ coefficient of 
resistivity, measured at $B=0$, with increasing pressure (see  
Fig.~3, inset). 

Our resistivity curves as a function of pressure are very reminiscent
of those observed in CeRu$_2$Si$_2$, in which it was found that there
is scaling of the form
\begin{equation}
M(H,T,p)/\mu_B = \psi(H/H_s(p), T/T_s(p) ) \label{scaling_eqn}
\end{equation}
where $H_s(p)$ and $T_s(p)$ are scaling parameters 
\cite{Puech88,Mignot88}. 
We find similar scaling behaviour when we plot the relative change of
the magnetoresistance $\Delta\rho = \rho (B) - \rho (B=0)$, normalised
by $\Delta\rho (B_M)$, versus $B/B_{M\parallel}$ and $B/B_{M\perp}$
($B_{M\perp}$ corresponding to the average of $B_{M\perp 1}$ and
$B_{M\perp 2}$).
This quantity is seen to evolve in a universal manner, implying that
$\rho(B,T,p)$ scales with the critical field and may be described by a
function $\rho(B/B_c(p),T)$ above 2.5\,K.

It is surprising that there is no need for a scaling parameter
$T_s(p)$, i.e. $T_s(p)$, which would naturally be identified with
$T^*$, is in \SrTwoLayer\ independent of pressure.
Thus, if we consider a critical regime centered on $B_M$ as
illustrated for example in Perry et al.\ \cite{Perry}, the dominant
effect of pressure is to shift the critical regime along the $B$
axis, without a significant change of the temperature dependence.

In the standard model \cite{Millis02,GGL,MnSi,Yamada01}, pressure
dependence is built in via one parameter only, $\chi(T=0)$.
For the case of \SrTwoLayer\ both $B^*$ and $T^*$ are expected to be
equally strongly pressure dependent, in stark contrast with
experimental observations.
The actual strong pressure dependence of $B^*$ and the weak $p$
dependence of $T^*$ require instead an additional pressure dependence
of the other parameters, such as the mode-mode coupling term or the
frequency and momentum spread of the spin fluctuation spectrum.
The required cancellation of $p$-dependencies for $T^*$ would
be purely accidental.
The non-divergence of $\chi$ at the critical point, combined with the
pressure independence of $T^*$, suggests that the underlying physics
of the metamagnetic transition in \SrTwoLayer\ is beyond the standard
sixth-order Ginzburg-Landau theory in which $M$ is the expansion
parameter \cite{Yamada93}.

CeRu$_2$Si$_2$ might provide an interesting alternative model: there
the metamagnetic-like transition is a crossover associated with the
destruction of antiferromagnetic correlations \cite{RossatMignod88};
moreover magnetovolume effects provide positive feedback to
dramatically sharpen what would, at constant volume, be a rather broad
crossover \cite{Puech88,Lacerda89}.  
If the magnetovolume coupling were just a little bit stronger it
appears that the crossover could become first order.

Although bulk thermodynamic measurements show that \SrTwoLayer\ is
nearly ferromagnetic \cite{Ikeda00}, recent neutron scattering
experiments have found that a fall in $\chi(T)$ below 16\,K at $B=0$T is
associated with the growth of antiferromagnetic spin fluctuations
\cite{Capogna02}.
Moreover lattice distortions in the form of tilts and rotations of the
oxygen octahedra couple strongly to magnetism in the ruthenates
\cite{Shaked00,Leitus99,Friedt01}, and they are thought to be decisive
in stabilising ferro- vs.\ antiferromagnetism \cite{Mazin97,Fang01}.
\SrTwoLayer\ has a rotation-distortion of $\sim7^{\rm o}$ about the
$c$-axis \cite{Shaked00,Inoue97} that provides an obvious mechanism
for positive magnetoelastic feedback for the metamagnetic transition.
So it seems plausible that a drop in antiferromagnetic correlations,
made discontinuous by positive feedback from magnetoelastic coupling,
could produce the observed first-order metamagnetic behaviour in
\SrTwoLayer.

This model might explain why $\chi$ does not diverge at the
metamagnetic critical point, and critical fluctuations at finite $q$ would
contribute to the large peak in the resistivity seen at $B_M$ in our 
resistivity curves.
It is not clear that the CeRu$_2$Si$_2$ model explains the weak
pressure dependence of $T^*$, but some experimental tests suggest
themselves: in particular, neutron scattering could be used to track
the field dependence of both antiferromagnetic correlations (as was
done in CeRu$_2$Si$_2$ \cite{RossatMignod88}) and the rotation angle
of the octahedra through the metamagnetic transition.

In conclusion, we have found a metamagnetic critical end-point at finite 
temperature for fields applied in the basal plane $B \perp c$ of \SrTwoLayer, 
though the peak in the susceptibility at the critical point is much 
smaller than expected. 
We have in addition shown that the application of pressure drives the critical 
point up in field, but surprisingly does not produce a substantial 
shift of the critical temperature. 

We wish to thank A.J. Millis, D. Forsythe, S.-I. Ikeda, A. Rosch, S.S.
Saxena, M. Uhlarz and H. v. L\"ohneysen.  We gratefully acknowledge
financial support from the ESF (FERLIN), NSERC (Canada), EPSRC (UK),
the Royal Society and the Leverhulme Trust.

\begin{figure}
\includegraphics[scale=0.5]{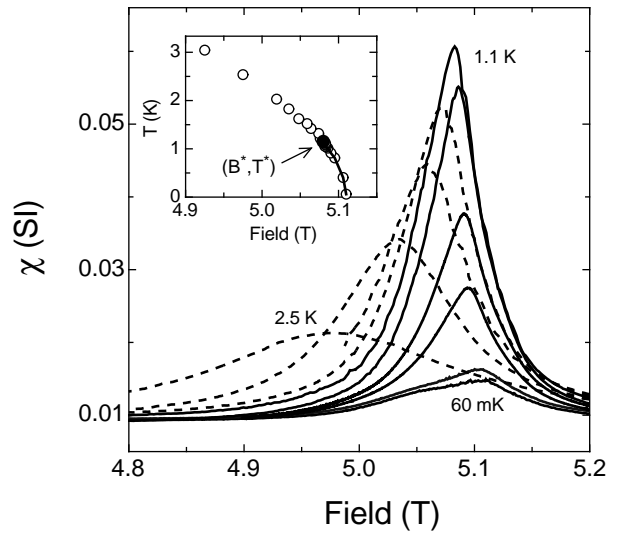}
\caption{A.c. susceptibility as a function of magnetic field for  
  $T = (0.06,0.4,0.8,0.9,1.0,1.1,1.3,1.5,1.8,2.5)$\,K .  The maximum
  peak corresponds to the critical point at
  $(B^{*},T^{*}$)=(5.08T,1.1K).  At the lowest temperatures (60\,mK
  and 400\,mK), the cusp-like features indicate first-order behaviour
  in $M$.  Hence for $T<T^*$ we have used solid lines to represent
  first-order behaviour; dashed lines represent crossover behaviour
  above $T^*$.  Inset: Temperature of the susceptibility peak as a
  function of magnetic field, with the solid line showing the line of
  first-order transitions, terminating in a metamagnetic critical
  end-point $(B^*,T^*)$. }
\end{figure}

\begin{figure}
\includegraphics*[scale=0.5]{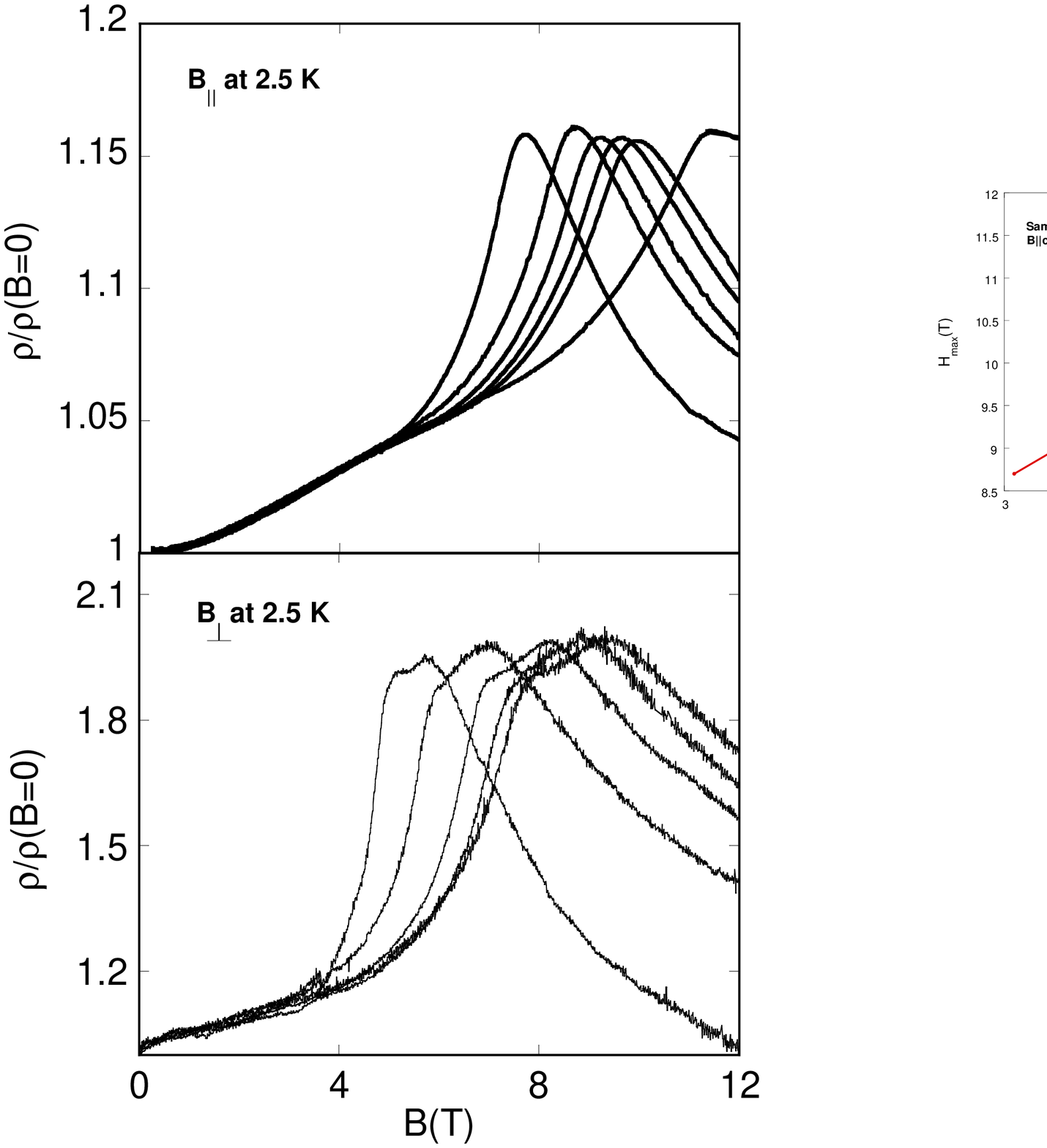}
\caption{Normalised magnetoresistance for $B~{||}~c$ (upper) and $B\perp c$ 
(lower) in Sr$_3$Ru$_2$O$_7$ for pressures up to 10\,kbar (increasing
pressure from left to right) at 2.5\,K.}
\end{figure}

\begin{figure}
\includegraphics[scale=0.4]{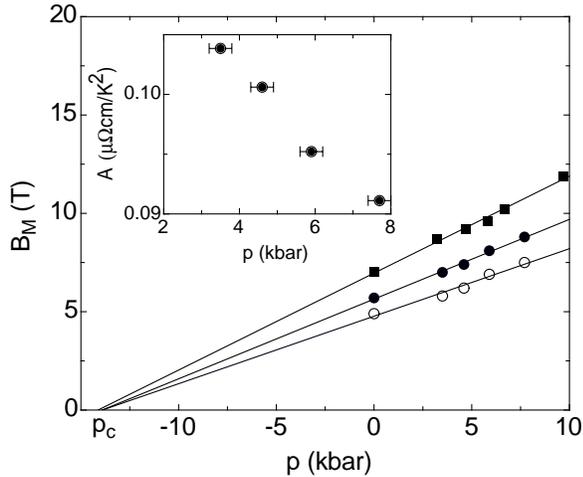} 
\caption{Increase of the metamagnetic transition
fields $B_{||}$, $B_{\perp 1}$ and $B_{\perp 2}$ in Sr$_3$Ru$_2$O$_7$
as a function of pressure, where $B_{\perp 1}$ corresponds to the
lower field value of the double transition and $B_{\perp 2}$ the
higher.  In any case, all three critical fields extrapolate to a
unique negative critical pressure $p_c \approx -14$\,kbar.  }
\end{figure}

\begin{figure}
\includegraphics[scale=0.37]{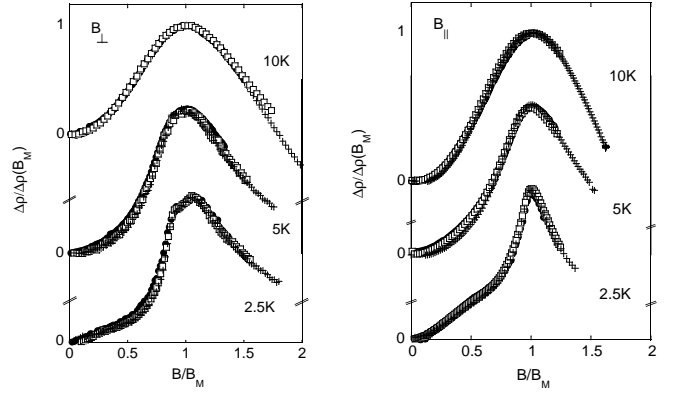}
\caption{Normalised magnetoresistance (refer to text) as a function of
the magnetic field scaled by the respective critical field $B_M$.  
Left: $B_M = (B_{\perp 1} + B_{\perp 2})/2$ is an average of the two 
critical fields for $B_{\perp c}$.  Right: $B_{M}=B_{||}$.}
\end{figure}

\end{document}